# REBCO-based dipole optimized for an ultimate-energy hadron collider

P.M. McIntyre *IEEE*, J.S. Rogers, and A. Sattarov

*Abstract*— REBCO tape has remarkable performance for super-conducting technology, but it is extremely expensive and it is a strongly anisotropic superconductor. The design of a 3.5 T collider dipole is presented in which all turns of the body winding are oriented so that the all turns of REBCO tape operate with maximum current density. Each turn within the winding contains a stack of Cu-clad REBCO tapes; all tapes within the turn are compressed to provide low-resistance Cu-Cu contact, but the successive turns are electrically isolated. Each turn is oriented so that the local magnetic field at the tapes is closely parallel to the tape surface. Sextupole is controlled using a current-programmed turn to provide excellent homogeneity over a 20:1 dynamic range. Issues of current re-distribution during ramping, quench stability, AC losses, synchrotron radiation, and cryogenics have been considered, and appear to provide favorable properties for an ultimate-energy hadron collider – 500 TeV.

## I. INTRODUCTION

REBCO tape has remarkable properties for use in supercon-ducting magnets. It can operate with useful current density up to liquid nitrogen temperature, and can produce very high magnetic field at temperatures of 20-40 K. The manufactured tape is ready to use as supplied, and does not require a final heat treatment after winding into a magnet winding – an important benefit compared to Nb$_3$Sn and Bi-2212.

But REBCO is extremely expensive, typically ~$90/m for a 6 mm wide tape capable of ~1000 A at 25 K. Also REBCO is a strongly anisotropic superconductor. Fig. 1 shows the de-pendence of the critical current $I_c$ in a 4 mm wide REBCO tape on the angle θ between the magnetic field $\vec{B}$ and the normal to the tape face when the tape is operating at 30 K temperature. The critical current $I_\parallel$ when $\vec{B}$ is oriented parallel to the tape surface (θ = 90°) is ~3 times greater than the critical current $I_\perp$ when $\vec{B}$ is oriented normal to the tape surface (θ = 0°). The tape must be oriented so that its face is oriented no more than Δθ~8° from the direction of $\vec{B}$ at the tape location (the green zone in the figure), if it is to be capable of operating with $I_c > 0.8$ $I_\parallel$. If one could orient the tapes of a winding so that the tape faces were everywhere aligned with the field, this excellent

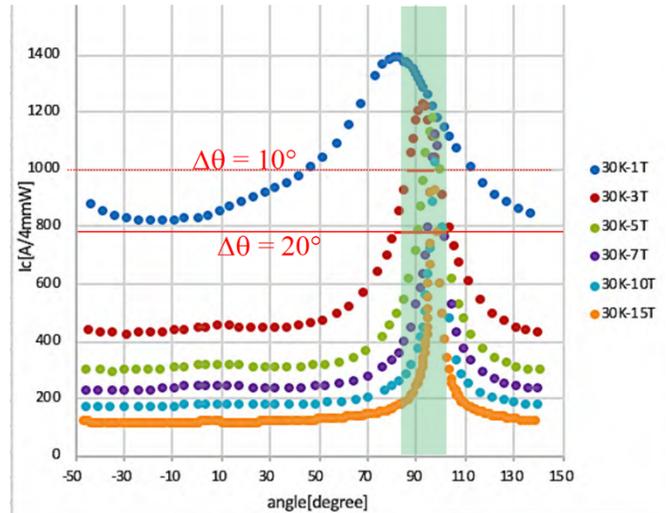

Fig. 1. Dependence of $I_c$ on the angle between the tape face normal and the magnetic field, for various field strengths (30 K temperature).

performance could be realized. But in a dipole winding $\vec{B}$ di-verges in the top/bottom regions of the body winding and ev-erywhere in the ends, so the winding current must be limited to ~$I_\perp$. In what follows we will examine a strategy by which to actually operate a REBCO dipole winding with a current ~0.8 $I_\parallel$. First, however, it is important to consider several aspects of current-sharing and inductive effects that complicate such a strategy.

Quench protection is important in an accelerator dipole. In the event that a quench occurs, the rapid change in winding cur-rent produces an inductive voltage among the turns of the wind-ing. Since inductance is proportional to the square of the num-ber turns, the inductive voltage should be limited by choosing a smaller number of turns (and correspondingly larger winding current). But since a single REBCO tape can carry only <1 kA even in when it is oriented so that its face is parallel to $\vec{B}$, it is necessary to assemble 10 or more tapes as a cluster for the con-ductor of a low-inductance dipole winding.

Clustering REBCO tapes has been used by several authors, including Roebel cable , Rutherford cable, CORC [1], cable-in-conduit [2], and stacked tapes [3]. In most cases REBCO tapes are stacked in a face-on cluster and then transposed by twisting

P.M. McIntyre is with the Accelerator Technology Lab, Texas A&M Univer-sity, College Station, TX 77845 USA and also with Accelerator Technology Corp. (e-mail: p-mcintyre@tamu.edu).

J.S. Rogers is with Accelerator Technology Lab, Texas A&M University, Cola

A. Sattarov is with Commonweath Fusion Systems, Boston, MA USA (e-mail: akhdiyorsattarov@gmail.com)

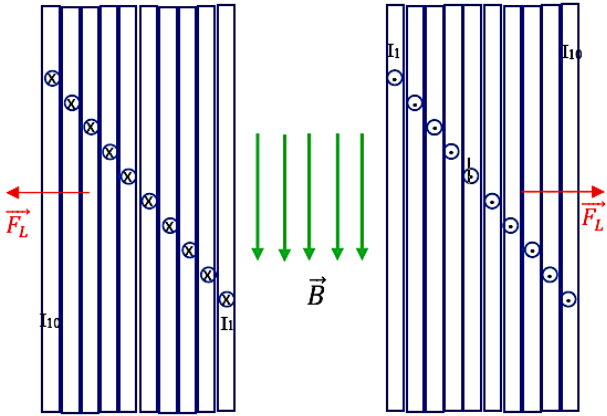

Fig. 2. Schematic model of a single-turn dipole winding of a 10-tape cable.

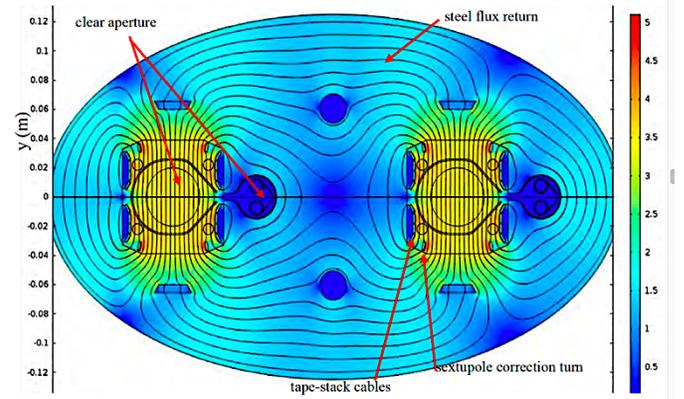

Fig. 3. 4 T dual dipole, in which the winding of each bore comprises 10 turns of REBCO tape cluster cable.

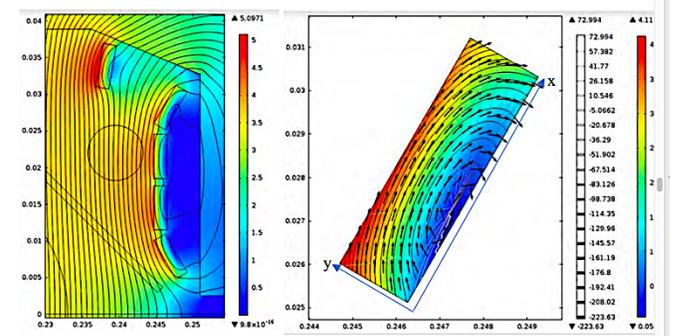

Fig. 4. Detailed view of the conformal winding of the dual dipole, showing the orientation of each tape cluster in the magnetic field at its location.

the overall cluster. The property of transposition is used in many superconducting winding designs to eliminate inductive coupling among the tapes of a cluster in each turn of the winding. To understand this effect, consider a cluster of tapes, in face-face resistive contact with one another within the cluster, carrying an overall current I that is distributed among the tapes as shown in Fig. 2.

When current is made to flow within a segment of length $\vec{L}$ in the winding, a Lorentz force $\vec{F}$ acts upon the current in a direction to shift the current outwards from the vertical mid-plane of the winding. The current $I_1$ in the innermost tape would experience an outward-directed force $\vec{F}_1 = I_1 \vec{L} \times \vec{B}$ that would transfer that current to the next tape (through the resistive copper cladding), and so forth, so that current concentrate in the outermost tape. As cable current were increased, the current $I_{10}$ in the outermost tape would reach its critical current limit prematurely and quench before significant current was flowing in the inner tapes.

Inductive forces on the currents among the tapes of a cable also produce AC losses as cable current is ramped up or down. If a tape is oriented normal to the magnetic field at the tape, increasing or decreasing the magnetic field induces an electromotive force $\mathcal{E}$ that would drive a loop current within the plane of the tape, reducing the tape's superconducting current capacity. Inductive forces on the currents in neighboring tapes likewise produce $\mathcal{E}$ between neighboring tapes that drives current through the normal-conducting copper layers between them.

In the above-mentioned uses of transposition in REBCO-based windings, the cluster of tapes is twisted along the length of the winding, so that each tape transposes from an inside location to an outside location as it traverses each twist pitch in the winding, so the inductive effects are mitigated. Transposition has the consequence, however, that the current in every tape is limited to $\sim I_\perp$.

## II. Conformal windings

A conformal winding is made by orienting the tapes in a tape-cluster winding so that all tape faces within each tape cluster are closely parallel to $\vec{B}$ at its location. The tape cluster is not transposed – the orientation of each tape in its local magnetic field is sustained in the parallel orientation that yields maximum superconducting current capacity.

Consider the 4 T dual dipole shown in Fig. 3, which is designed as a sector dipole for an ultimate-energy (500 TeV) hadron collider to be housed in a 2000 km circular pipeline supported in neutral buoyancy at a depth of 100 m in the Gulf of Mexico [4]. The winding of each bore comprises a single layer of 10 turns of REBCO tape-cluster cable.

Fig. 4 shows a detailed cross-section of the dipole winding and shows how each turn is oriented so that its face is parallel to its local magnetic field. The current capacity of the $n$th tape in each cluster can be estimated by extracting the local sheet current density $K_n(x)$, adding it up for the entire tape width: $I_{cn} = \int_0^w K_n(x)\,dx$. Using this method, for the particular dipole design shown, the cable critical current is $\sim 80\%\ I_{\parallel}$. The conformal winding thus requires three times less REBCO tape as would a winding in which all tape clusters were oriented vertically, or one utilizing a transposed cable as in prior art.

Two versions of the design have been prepared and optimized: one with $N$=18 tapes in the tape-stack cable to produce 3.5 T bore field at short-sample limit; the other with $N$=28 tapes in the tape-stack cable to produce 4.5 T bore field. Both versions have been optimized to produce approximately homogeneous field distribution in the bore.

## III. Current-sharing in a tape-stack cable

As the dipole is ramped to increase the bore field from its value $B_i$ at injection energy to $B_c$ at collision energy, the current within each tape-cluster cable is increased proportionately. As in the preceding discussion, the magnetic field at each cable produces a Lorentz force $\vec{F}$ that pushes current flowing within each tape-cluster cable away from the dipole bore. Thus, even if current were injected so that it was initially distributed equally among all tapes within a first turn of cable, the Lorentz force would re-distribute current to the outermost tapes within the winding. It would therefore seem that as coil current were increased the outermost tape would quench when the overall cable current was still only a fraction of the desired current!

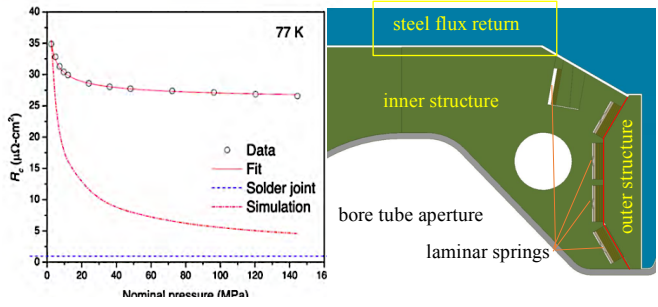

Fig. 5. a) Contact resistance between two copper-clad REBCO tapes as a function of compression; b) structural assembly of a conformal winding showing the support of each tape-stack cable and compression by a laminar spring.

But REBCO can operate at 30 K, where the heat capacity of the tape ($\sim T^3$), and the conduction to remove heat $\sim (T_{hot}-T)$, are both much greater, so it is possible to operate a cable of stacked non-insulated tapes *without transposition* and rely upon the 'soft' approach to quench in each tape to force re-distribution of current within the cable as the cable current is further increased. This strategy has been used to good effect in 'no-insulator' (NI) pancake windings for high-field solenoids [5]. We now analyze the dynamics of quasi-equilibrium current-sharing among the $\sim 10$ tapes within a non-transposed cable, the approach of Noguchi [6] [7].

The REBCO layer within each exhibits a retarding electric field E that is current dependent: $E_z = E_0(I/I_c)^n$ (1) where $E_0 = 10^{-6}$ V/m is the quench criterion, $I_c$ is the quench current for the conditions $(B,T,\theta)$ for that tape, and $n=24$ is the index that characterizes the power-law dependence of the superconductor-normal transition for REBCO.

The dynamics is analogous to the Hall effect in which the superconducting transport is acted upon by the transvers Lorentz force, and by a transverse electric field produced by the potential difference between neighboring tapes when they carry different currents, and by a contact resistance $R_c$ between adjacent tapes through which current is displaced by Lorentz forces.

Following Ref. 6, the time dependent distribution of current in a tape-stack cable can be estimated in a simple model in which the full length of one half-turn of the tape-stack cable is treated as a series-parallel L/R network. Each tape within a half-turn of one cable has a self-inductance per unit length

$$\tilde{L} = \frac{\mu_0 x}{wg} = 4 \times 10^{-4} H/m \quad (2)$$

and a power-law series resistance

$$\tilde{R}_s = \frac{E}{I} = \frac{E_0}{I_0}\left(\frac{I}{I_0}\right)^{n-1} = (0.7 n\Omega/m)\left(\frac{I}{I_0}\right)^{n-1} \quad (3)$$

where w=6 mm is the tape width, g=10 cm is the vertical gap in the steel flux return, and x~10 cm is the horizontal width of the tape loop.

Lu *et al.* [8] measured the dependence of the contact resistance $R_c$ upon the compression among the tapes in the stack, shown in Fig. 5a.

As shown in Fig. 5b, each turn of tape-cluster cable in the conformal winding is supported by a laminar spring that provides ~1 MPa uniform compression within the tape-cluster, corresponding to contact resistance $R_c \sim 35$ $\mu\Omega$-cm². The parallel resistance of a length $\mathcal{L}$ of a tape to each of its neighbors is

$$R_p = \frac{R_c}{\mathcal{L}w} = \frac{(0.6\mu\Omega \cdot m)}{L} \quad (4)$$

From these quantities, we can extract two results that characterize the scale of current-sharing. First, the scale length $\lambda$ over which this homogenization operates is the winding length for which $R_p \sim R_s$:

$$\lambda = \sqrt{\frac{R_c}{wR_s}} = 29m\left(\frac{I}{I_c}\right)^{-11.5} \quad (5)$$

$\lambda$ is much longer than any reasonable winding length, so the current distribution will relax uniformly along the winding.

Second, we can estimate the time constant with which a difference in current between successive tapes in a tape-stack cable relaxes to an equilibrium governed by the Lorentz force and the 2-D distribution of resistance within a tape-stack cable. The change in inductance along one winding length $\mathcal{L}$ between one tape and the next is

$$dL = \frac{\mu_0}{wg}dx = 0.4\mathcal{L} \ \mu H. \quad (6)$$

Current-sharing poses a further challenge for field homogeneity, however. At injection field, the current in each tape cluster is located mainly n the outermost tape; at collision field, the current is ~equally shared, so the 'current position' for that cable turn is shifted inwards by half the tape width. The multipoles have been evaluated for the magnetic design of Figure 5 for these two limiting cases. The difference in the calculated multipoles is $\Delta b_n < 0.5 \times 10^{-4}$ for all multipoles! This result might seem remarkable, but it is actually a consequence of the conformal design strategy: because each tape cluster is oriented so that the tape faces are closely parallel to the field at conductor, the field distribution is insensitive to the horizontal position of the 'current center position' of that cluster.

## IV. SUPPRESSION OF PERSISTENT-CURRENT MULTIPOLES

In the conformal winding, all tape-stack cables are oriented so that tape faces are closely parallel to the magnetic field at the tape. As the winding current is ramped up or down, there is no induced emf in the plane of any tape, so persistent-current loops are strongly suppressed. This is a unique property of the conformal winding of tape-stacks, not true for any other geometry of wire or cable. The multipoles from persistent-currents are a significant challenge for beam dynamics at injection field for colliders, and conformal windings naturally suppress them.

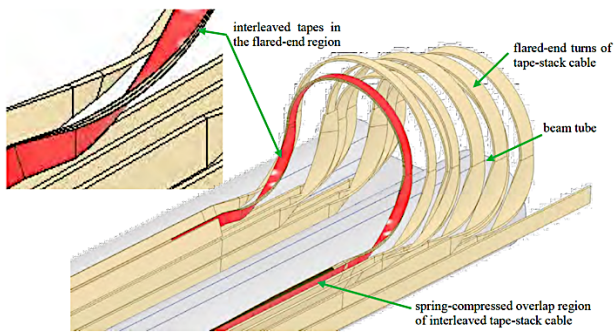

Fig. 6. Schematic view of a reinforced end winding, in which additional tape segments are interleaved between the tapes of each tape-stack cable turn..

## V. END WINDING

At each end of a dipole winding, the magnetic field flares outward both vertically and horizontally and returns to the surrounding steel flux return, as shown in Fig. 6. In the end region of the dipole, each turn of tape-stack cable must be formed along a catenary curve that connects one turn of the body winding from one side to the other and flares out of the midplane to provide clearance for the beam tube of the dipole.

Provision is made to interweave additional tape segments between the tapes within the flared-end region of each tape-stack cable. The reinforced region can accommodate current transfer among the tapes of the tape-stack cable and the reinforcing tapes to provide twice the current-carrying capacity in the end winding region.

## VI. .CONFORMAL SUB-WINDINGS FOR A HIGH-FIELD DIPOLE

Fig. 7 shows a 2-layer conformal sub-winding of REBCO tape-cluster in an 18 T hybrid dual dipole designed for the requirements of a 100 TeV hadron collider in a 100 km tunnel. The dipole was originally designed using a Nb₃Sn outer sub-winding and a Bi-2212 inner sub-winding (shown in left half of the quadrant), each composed of SuperCIC round cable [9]. The right half of the quadrant shows replacement of the Bi-2212 winding by a conformal REBCO tape-stack winding. The field homogeneity was preserved, the quantity of superconductor was reduced by half (although it is still prohibitively expensive!). By the above arguments, persistent-current multipoles from the REBCO sub-winding should be significantly reduced by the conformal geometry. This final example illustrates the power of the conformal winding strategy to get the maximum performance from a REBCO winding, and correspondingly to reduce the quantity of expensive superconductor for a given application.

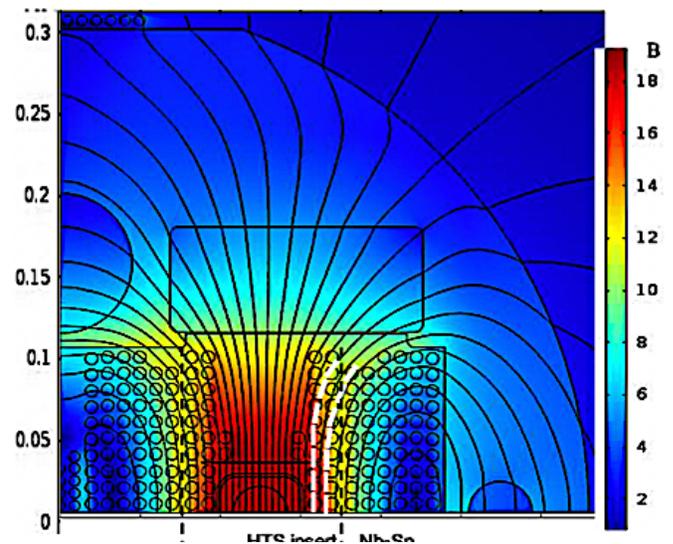

Fig. 7. 18 T hybrid dual dipole for a 100 TeV hadron collider. The winding comprises an outer Nb₃Sn sub-winding and an HTS inner sub-winding. The left half shows the inner sub-winding using Bi-2212 CIC; the right half shows the inner sub-winding using conformal REBCO tape-cluster cable.